%% file: multispectral_compression_hevc.tex
\documentclass[conference]{IEEEtran}
\IEEEoverridecommandlockouts
\IEEEpubid{\makebox[\columnwidth]{978-1-7281-9320-5/20/\$31.00 \copyright 2020 IEEE \hfill} \hspace{\columnsep}\makebox[\columnwidth]{ }}

\usepackage{cite}
\usepackage{amsmath,amssymb,amsfonts}

\usepackage{graphicx}
\usepackage{textcomp}
\usepackage[table,xcdraw]{xcolor}
\usepackage{tikz}
\usepackage{pgfplots}
\usepackage{subfig}
\usepackage{bm}
\usepackage{multirow}
\usepackage{hhline}
\usepackage[linesnumbered,ruled,vlined]{algorithm2e}

\usepackage[firstpage=true]{background}
\usepackage{hyperref}
\SetBgContents{%
	\fbox{%
		\parbox{\textwidth}{%
			\footnotesize © 2020 IEEE. Personal use of this material is permitted. Permission from IEEE must be
			obtained for all other uses, in any current or future media, including
			reprinting/republishing this material for advertising or promotional purposes, creating new
			collective works, for resale or redistribution to servers or lists, or reuse of any copyrighted
			component of this work in other works. DOI: \href{https://doi.org/10.1109/MMSP48831.2020.9287132}{10.1109/MMSP48831.2020.9287132.}}}%
}	
\SetBgScale{1}
\SetBgAngle{0}
\SetBgPosition{current page.north}
\SetBgVshift{-1.6cm}
\SetBgColor{black}
\SetBgOpacity{1}

\def\BibTeX{{\rm B\kern-.05em{\sc i\kern-.025em b}\kern-.08em
    T\kern-.1667em\lower.7ex\hbox{E}\kern-.125emX}}
\begin{document}

\title{\vspace{0.5em} Multispectral Image Compression Based on HEVC Using Pel-Recursive Inter-Band Prediction
}
\author{\IEEEauthorblockN{ Anna Meyer, Nils Genser, and André Kaup}
\IEEEauthorblockA{\textit{Multimedia Communications and Signal Processing} \\
\textit{Friedrich-Alexander-Universität Erlangen-Nürnberg (FAU)}\\
Erlangen, Germany \\
\{ anna.meyer, nils.genser, andre.kaup \}@fau.de}

\vspace{-1.5\baselineskip}}
\maketitle
\begin{abstract}
Recent developments in optical sensors enable a wide range of applications for multispectral imaging, e.g., in surveillance, optical sorting, and life-science instrumentation. Increasing spatial and spectral resolution allows creating higher quality products, however, it poses challenges in handling such large amounts of data. Consequently, specialized compression techniques for multispectral images are required.
High Efficiency Video Coding (HEVC) is known to be the state of the art in efficiency for both video coding and still image coding. In this paper, we propose a cross-spectral compression scheme for efficiently coding multispectral data based on HEVC. Extending intra picture prediction by a novel inter-band predictor, spectral as well as spatial redundancies can be effectively exploited. Dependencies among the current band and further spectral references are considered jointly by adaptive linear regression modeling. The proposed backward prediction scheme does not require additional side information for decoding.
We show that our novel approach is able to outperform state-of-the-art lossy compression techniques in terms of rate-distortion performance. On different data sets, average Bj{\o}ntegaard delta rate savings of 82~\% and 55~\% compared to HEVC and a reference method from literature are achieved, respectively.
\end{abstract}

\begin{IEEEkeywords}
Multispectral imaging, lossy compression, predictive coding, High Efficiency Video Coding (HEVC).
\end{IEEEkeywords}

\section{Introduction}
Multispectral images capture the same spatial region multiple times, each time at another optical wavelength. These measurements of different frequencies, typically beyond the visible spectrum, make the image data three-dimensional~\cite{tretter05}. The most common type of multispectral images are RGB color images, which are often extended by Ultraviolet (UV) and Near-Infrared (NIR) data. Hyperspectral images, however, usually consist of dozens of narrow or even contiguous spectral bands.
With developments in image sensors and electronics, multispectral cameras are becoming more affordable. Thus, new applications emerge: multispectral imaging systems for airborne agricultural product management \cite{huang10}, object detection in traffic \cite{takumi17} or monitoring health via mobile phones \cite{Kim16}. 
\begin{figure}[tbp]
	\centering
	\includegraphics[width=0.495\textwidth]{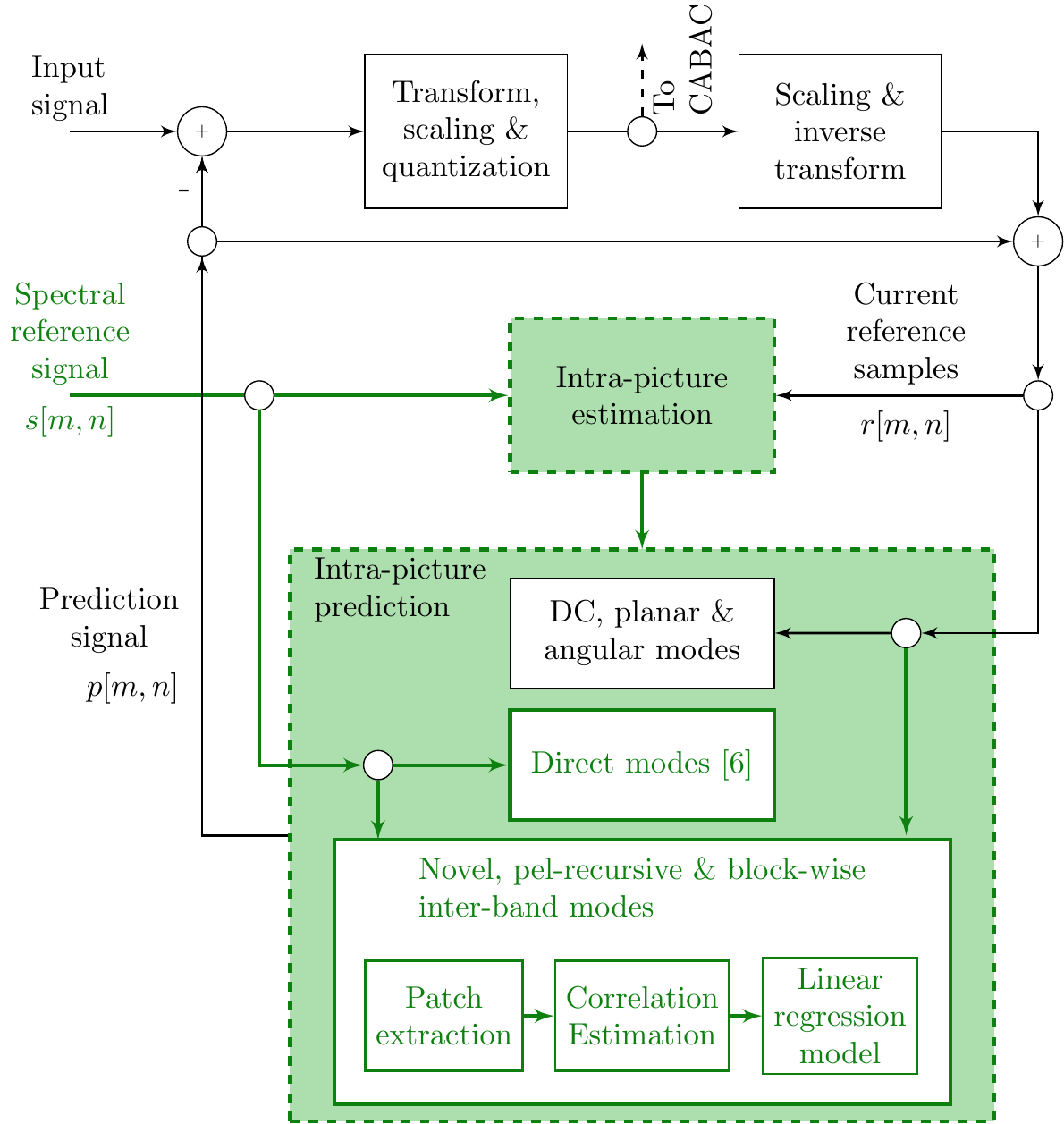}
	\caption{HEVC intra prediction with modified elements highlighted in green. The proposed method adds five inter-band prediction modes, which rely on an additional 3-channel spectral reference signal, to the existing intra modes. The three direct modes are described in Sec.~\ref{sec:direct}. The pel-recursive and block-wise mode differ in how the depicted steps for estimating prediction parameters are applied. The former is further explained in Sec.~\ref{sec:pixel-wise} and the latter in Sec.~\ref{sec:block-wise}.}
	\label{fig0}
\end{figure}

Due to the higher amounts of data resulting from the third, i.e., spectral image dimension, specialized and efficient compression techniques are required.
A multispectral coding scheme needs to be adapted to distinct features of multispectral data to achieve a higher compression ratio. Hence, it needs to exploit spectral in addition to spatial redundancies. While spatial redundancy is due to dependencies among neighboring pixels in a single band, correlations between co-located pixels of different bands cause spectral redundancies. As the bands capture the same scene, their edge information is typically highly similar, whereas the dynamic ranges of pixel intensities differ \cite{Salamati12, SPIE14}.
In this paper a compression algorithm is introduced, which is able to adaptively exploit these similar edge information across spectral bands. The proposed method is based on the High Efficiency Video Coding (HEVC) standard. Fig.~\ref{fig0} provides an overview of the modified elements indicated in green: A novel cross-spectral predictor extends intra-picture prediction without requiring additional side information for decoding. Pure HEVC intra coding has been shown to outperform still image codecs \cite{nguyen15}, making HEVC a benchmark in image as well as video coding.

The paper is organized as follows. In Sec.~\ref{sec2}, state-of-the art compression techniques are reviewed. Then, the novel method is presented in Sec.~\ref{sec3} and evaluated for multiple data sets in Sec.~\ref{sec4}. Finally, Sec.~\ref{sec5} concludes the paper.
\section{State of the Art}\label{sec2}
In general, three main types of multispectral compression techniques can be named. 
Vector quantization based approaches interpret the pixel values of different bands at the same spatial position as vectors and directly quantize the data using trained codebooks \cite{ryan97}. Prediction-based techniques decorrelate the image data utilizing spatial, spectral or hybrid predictors \cite{Lin10}. The third main type achieves  decorrelation by applying transforms. Commonly used transforms include the discrete cosine transform \cite{Salamati12}, discrete wavelet transform \cite{tang03}, or  Karhunen-Lo\`{e}ve Transform \cite{wang09}.

The Consultative Committee for Space Data Systems recommended a lossless as well as near-lossless compression standard consisting of a predictor followed by an entropy encoder \cite{CCSD}. The value of each image sample is predicted based on the nearby samples in a small three-dimensional neighborhood using an adaptively weighted prediction algorithm. The standard is extended by a spectral preprocessing transform \cite{CCSD_spectral}. It aims at generating a transformed image that is compressed more efficiently by a two-dimensional encoder. This principle has been widely used in the literature, especially in combination with the image codec JPEG 2000 \cite{Du07, penna07, B_scones_2018}.

However, in \cite{Dusselaar15} and \cite{Paul16} the authors argue that a shift is required from the original pixel intensity based traditional image codecs to residual based video coders for hyperspectral compression. They modified inter prediction in HEVC by adding a spectral reference image and treating each band as a video frame. The additional reference is an instant spectral band generated using Gaussian Mixture-based Modeling. 
Building on standards like HEVC provides the benefit of being able to directly utilize its further developments for multispectral compression.
In~\cite{SPIE14}, Gao et al. extend HEVC intra prediction by two inter-band modes that use a previously decoded spectral band as additional reference. The band to be coded is predicted from spectral co-located blocks: The residual is either directly coded or based on a linear relationship. In the latter case, the prediction parameters obtained by least mean square optimization need to be coded for every Prediction Block~(PB), which causes considerable side information to be transmitted.

To overcome this drawback, we propose a novel backward prediction scheme, which extends intra prediction in HEVC based on linear regression modeling. The main differences to Gao et al. \cite{SPIE14} are: (i) the prediction relies on adjacent as well as previously coded samples instead of the original signal, (ii) three spectral bands are used as reference signal rather than a single band, (iii) pixel-wise regression problems can be solved in contrast to a block-wise problem, and (iv) a processing order must be formulated. Consequently, blocking artifacts are prevented and the modeling error is minimized over a smaller region, making the prediction highly local. We interpret the prediction problem as extrapolation problem without side information, i.e., no parameters need to be transferred. This results in significant rate savings on a variety of image contents, as will be shown in Sec.~\ref{sec4}.
\section{Proposed Method}\label{sec3}
In the following, the Pel-Recursive Inter-Band Prediction (PRIBP) algorithm is introduced. Direct and regression based inter-band prediction modes extend intra prediction in HEVC, as illustrated in Fig.~\ref{fig0}. Both en- and decoder are affected.
\subsection{Intra-Picture Estimation}
The decision whether to apply intra or spectral prediction is made at prediction unit (PU) Level. For each PU, a candidate list of possible modes is generated. Integrating inter-band into intra picture prediction, the number of available intra modes increases from 35 to 40 with three direct and two causal inter-band modes. Afterwards, the mode corresponding to the smallest rate-distortion (RD) cost is chosen. Due to the fact that inter-band correlations vary significantly in different image regions~\cite{tretter05}, it is desirable to adaptively switch between spectral and intra prediction. In our experiments (see Sec.~\ref{sec4}), the new inter-band modes were chosen for more than 78~\% of PUs on average. This implies that the required rate for intra mode coding is not affected significantly, since the inter-band modes are among the most probable modes frequently.
\subsection{Block Size for Inter-Band Prediction}
If inter-band prediction was selected for a PU, the prediction signal is computed on Transform Unit (TU) instead of PU level. Hence, a possible TU quadtree split is always considered. A TU tree structure has its root at the Coding Unit (CU) level \cite{sullivan}. Transform block (TB) sizes range from $4\times4$ to $32\times32$, resulting in a finer resolution compared to using PBs with a maximum size of $64\times64$. Therefore, more reference samples of neighboring blocks are available for each inter-band prediction problem. 
\subsection{Direct Inter-Band Prediction} \label{sec:direct}
For each of the three direct modes, a co-located block from one of the three channels of the spectral reference signal is directly used as prediction signal. The remaining two inter-band prediction modes, which rely on boundary samples of neighboring blocks, will be described in the following.
\subsection{Pel-Recursive Inter-Band Prediction} \label{sec:pixel-wise}
\subsubsection{General structure and involved signals}
The prediction signal $p[m, n]$ is calculated pixel-wise for each TB of size $N\times N$, depicted by the integer coordinates $m$ and $n$. 
The prediction incorporates the information available in two signals: (i) a spectral reference $s[m, n]$ and (ii) a reference $r[m, n]$ of the current band to be coded. Both signals can be obtained at the encoder as well as at the decoder. Consequently, no parameters have to be transferred and the prediction problem is considered as prediction problem without side information.
\begin{figure}[tbp]
	\centering
	\includegraphics[width=.43\textwidth]{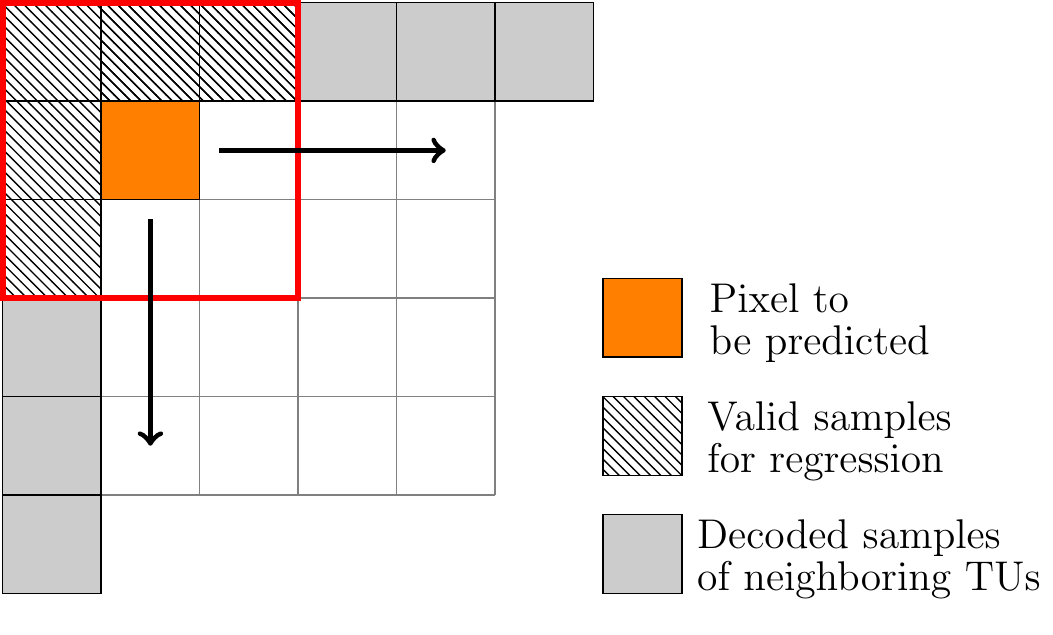}
	\caption{Example of a $3\times3$ patch in red. The pixel to be predicted shown in orange is the center of the patch. The arrows indicate that pixels only depending on causal samples of neighboring TBs are predicted first.}
	\label{fig1}
\end{figure}
The reference $r[m, n]$ of the current band consists of two types of valid support pixels:
(i) Previously predicted pixels inside the currently coded TB and, (ii) reconstructed pixels extracted from adjacent, previously decoded TBs. Fig.~\ref{fig1} illustrates the location of these $2N+3$ pixels in gray.
Due to the coding structure of HEVC, upper and left adjacent PBs and TBs are frequently available \cite{Lainema12}. If neighboring pixels are unavailable, e.g. at picture or slice boundaries, the missing sample values are filled with the closest available boundary samples as specified in \cite{Lainema12}.

Fig.~\ref{fig0} and the function "predict" in line 6 of Algorithm~\ref{alg1} give an overview of the steps involved in the computation of each set of prediction parameters. As opposed to the block-wise mode (see Sec.~\ref{sec:block-wise}), parameter estimation is performed for every prediction value. First, one of three previously decoded bands is chosen as the spectral reference $s[m, n]$. For that, the Pearson correlation coefficient (PCC) is employed, which is defined as follows:
\begin{equation*} 
\rho(\bm{r}, \bm{s}) = \frac{\sum_{i=1}^{n} (r_i - \mathrm{E} \{\bm{r}\})(s_i - \mathrm{E} \{\bm{s}\})}{\sqrt{\sum_{i=1}^{n}(r_i  - \mathrm{E} \{\bm{r}\})^2} \sqrt{\sum_{i=1}^{n}(s_i - \mathrm{E}\{\bm{s}\})^2}}. \tag{1}
\end{equation*}
Secondly, a linear regression model is calculated to fit the data from the spectral reference $s[m, n]$ into the one from the current band's reference $r[m, n]$ \cite{Genser19}.
For these two processing steps, patches of size $b\times b$ are extracted from the two reference signals at the same position. Each patch is centered at the pixel to be predicted, as Fig.~\ref{fig1} shows. Only valid, i.e., already predicted pixels from $r[m, n]$ are taken into account. Thus, the respective samples of each reference signal are extracted and denoted as $\bm{r}_b$ and $\bm{s}_b$, with $b$ indicating the patch size.
\subsubsection{Spectral Reference Band Selection}
Using patches of size $5\times5$ for the computation of $\rho(\bm{r}_5,\bm{s}_5^i)$ with $i=1,2,3$, the band corresponding to the highest absolute value is selected as spectral reference. A higher absolute PCC value indicates a stronger linear correlation between the reference signals. Hence, the maximum value of one suggests a true linear relationship. In that case, the value to be coded can entirely be predicted by the available reference samples. Thus, the prediction error, i.e., the residual becomes zero, resulting in the highest possible rate reduction due to all correlations being removed.
\subsubsection{Linear Regression Model}
For each pixel, a linear regression problem is formulated as
\begin{equation*}
\bm{r}_3 \approx \alpha \cdot \bm{s}_3 + \beta \tag{2}
\end{equation*}
with slope and offset scalars $\alpha$ and $\beta$, respectively. Hence, $\alpha$ and $\beta$ must be determined to minimize the squared model error
\begin{equation*}
\mathop {\min }\limits_{\alpha,\beta} \left\| {\alpha \cdot {\bm{s}_3} + \beta - {\bm{r}_3}} \right\|_2^2. \tag{3}
\end{equation*}
The minimization problem is solved by applying a least squares fitting.
Finally, the prediction value is obtained as:
\begin{equation*}
p_{m, n}= \alpha \cdot s_{m, n} + \beta \tag{4}
\end{equation*}
with $m, n = 0, \hdots, N-1$. The prediction values are iteratively computed according to the processing order specified in Algorithm~\ref{alg1}, since previously predicted pixels are used as support. As indicated by the arrows in Fig.~\ref{fig1}, pixels only depending on boundary samples are predicted first. The remaining pixels are processed row-wise, resulting in a reasonable complexity and number of valid support pixels for each pixel to be predicted.
\begin{algorithm}[tb]
	\caption{Pel-recursive inter-band prediction with processing order for a $N \times N$ TB.} \label{alg1}
	\SetKwInput{KwInput}{Input}                
	\SetKwInput{KwOutput}{Output}              
	\DontPrintSemicolon
	
	\KwInput{$r[m,n]$ and $s[m, n]$}
	\KwOutput{$p[m,n]$}
	
	\SetKwFunction{FSub}{predict}
	
	predict($0,\;n$) $ \; \forall \: n \in [0,N)$ \\
	predict($m,\;0$) $ \; \forall \: m \in [1,N)$ \\
	\For{$m=1,  \hdots, N-1$} 
	{
	\For{$n=1,  \hdots, N-1$} {
	predict($m, \;n$)
}
}
	
	\SetKwProg{Fn}{Function}{:}{}
	\Fn{\FSub{$m$, $n$}}{
		$\bm{s}_5^1,\bm{s}_5^2, \bm{s}_5^3, \bm{r}_5$ = patchExtraction($m$,$n$) \\
		$\bm{s}_3, \bm{r}_3$ = correlationEstimation($\bm{s}_5^1,\bm{s}_5^2, \bm{s}_5^3, \bm{r}_5$) \\
		$\alpha$, $\beta$ = estimateParameters($\bm{s}_3, \bm{r}_3$) \\	
		\KwRet $p_{m,n} = \alpha \cdot s_{m, n} + \beta$
	  }

\end{algorithm}

Due to the small patch size and pixel-wise estimation, the prediction error is highly local. With the finest coding resolution possible in HEVC (TU level), this leads to adaptively processed edge information and thus to preserved details.
\subsection{Block-wise Inter-Band Prediction} \label{sec:block-wise}
The block-wise inter-band prediction mode is similar to the pel-recursive mode explained in Sec.~\ref{sec:pixel-wise}. However, correlation estimation and linear regression modeling is only performed once per block. Therefore, the same slope and offset scalars $\alpha$ and $\beta$ are used for the computation of every prediction value in $p[m,n]$. All $2N+3$ boundary pixels of neighboring TBs in $r[m,n]$ and the corresponding samples from $s[m,n]$ are used for the inter-band prediction problem.
\subsection{Coding order}
Adjusting the prediction structure based on the similarity of spectral bands leads to an increase in prediction accuracy. Consequently, the prediction error can be decreased, resulting in additional rate savings.
The proposed coding order can be employed: (i) if at least four bands are to be coded, and (ii) if coding dependencies between different bands can be tolerated.
The first three bands are coded independently, usually starting with RGB (if available). Afterwards, the remaining bands are sorted in ascending order according to their structural similarity index (SSIM) \cite{Wang04} with respect to one anchor band. After being coded, the least similar reference band is replaced by the previous band.
Therefore, the spectral references for each band are more similar to the band to be coded next. Hence, more correlations can possibly be exploited by sorting.    
\section{Experimental Results}\label{sec4}
The proposed PRIBP has been integrated in the HEVC test model (HM) 16.20 software. A reference implementation of our novel algorithm is available at:
\texttt{https://gitlab.lms.tf.fau.de/lms/pribp}

For comparing PRIBP with the state-of-the-art compression techniques HEVC Intra, JPEG 2000 and the HEVC-based multispectral approach by Gao et al. \cite{SPIE14}, experiments were conducted on different publicly available multi- and hyperspectral data sets. Gao et al. \cite{SPIE14} was chosen as reference method since it modifies intra prediction as well (see Sec.\ref{sec2}).

The first multispectral database provided by the Computer Vision Laboratory (CAVE) of Columbia University consists of 32 scenes. For each scene the $512\times512$ reflectance data of 31 bands ranging from 400 to 700 nm and a representative RGB image is available. The multispectral images capture a variety of objects and materials.
The second multispectral data set called \textit{Natural Scenes} shows urban scenes, grass, earth and trees \cite{Foster06}. The measured wavelengths range from 400 to 720 nm with a size of e.g. $1020 \times1339\times32$ for scene 5. Like in \cite{Dusselaar15} and \cite{Paul16}, scenes 5 and 6 were used for evaluation.
The Airborne Visible/Infrared Imaging Spectrometer (AVIRIS) collects hyperspectral images with 224 bands ranging from 400 to 2500 nm. The well-known Cuprite data set shows geological features of a mining district in Nevada. With the all zero bands excluded its reflectance data of size $ 512 \times 680 \times 206$ is used as third hyperspectral data set.

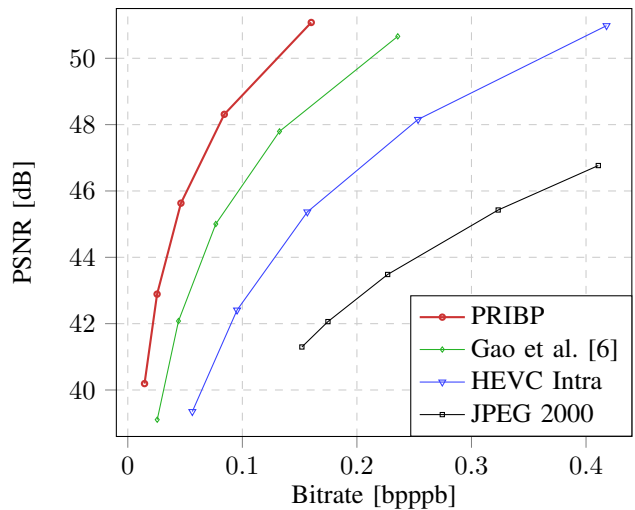
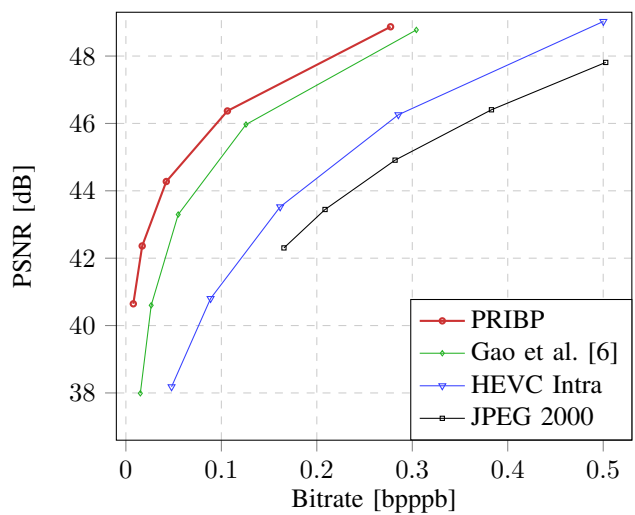
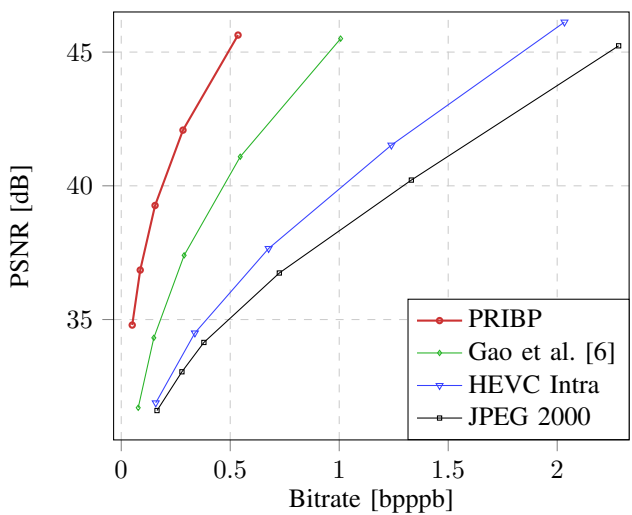
\begin{figure}[htbp]
	\subfloat[CAVE data set \label{1a}]{%
		\input{figures/cave_complete.tex}	  
	}
	\\
	\subfloat[Natural scenes (scene 5 and 6) \cite{Foster06} \label{1b}]{%
		\input{figures/scene5+6.tex}
	}
	\\
	\subfloat[Cuprite (AVIRIS) \label{1c}]{
		\input{figures/cuprite.tex}
	}
	\caption{Rate Distortion performance in terms of PSNR and bitrate. The three data sets are evaluated using the proposed PRIBP compared to HEVC Intra, the multispectral compression technique introduced by Gao et al. \cite{SPIE14} and JPEG 2000.}
		\label{fig3}
\end{figure}

The prediction parameters required for the pel-recursive mode (see Sec.~\ref{sec:pixel-wise}), i.e. the patch sizes for correlation and regression, were trained on a random subset of the CAVE data set consisting of 5 scenes. The patch sizes resulting in the best RD performance were determined to be 5 and 3 for correlation and regression, respectively. The remaining scenes were used for evaluation (see Fig.~\ref{1a}). Fig.~\ref{fig3} gives an overview of the RD performance of the proposed PRIBP in comparison to HEVC Intra, JPEG 2000 and Gao et al. \cite{SPIE14}. The compression performance is measured in terms of PSNR in dB and bitrate in bits per pixel per band (bpppb). For the HEVC-based methods Quantization Parameters (QPs) of 17, 22, 27, 32 and 37 are shown, including the values specified in the common HM test conditions \cite{qps}. The proposed method clearly outperforms the reference methods.

As JPEG 2000 is not competitive, the coding efficiency compared to the HEVC approaches is further evaluated in terms of Bj{\o}ntegaard delta rate (BD rate), which is computed using piece-wise cubic interpolation \cite{bdrate}. 
\bgroup
\def\arraystretch{1.4}
\begin{table}[]
	\caption{Performance of the proposed method PRIBP compared to HEVC Intra and Gao et al. \cite{SPIE14} in terms of BD rate.}
	\begin{center}
	\begin{tabular}{|c|c|c|c|}
		\cline{2-4} 
		\multicolumn{1}{c|}{} & \multirow{2}{*}{\textbf{Data set}}& \multicolumn{2}{c|}{\textbf{BD Rate savings {[}\%{]}}}                       \\ \cline{3-4} 
         \multicolumn{1}{c|}{} &           & QP = 0,7,12,17 & QP = 22,27,32,37\\ \hhline{|=|=|=|=|}
          
		 \multirow{3}{*}{\rotatebox[origin=c]{90}{\parbox[c]{1.2cm}{\centering \textbf{HEVC Intra}}}} 
		 & CAVE                            & 65.63 & 81.66     \\ \cline{2-4} 
		&Natural Scenes                   & 15.83  & 82.01       \\ \cline{2-4} 
		&Cuprite                 &   56.46     & 83.62  \\ \hhline{|=|=|=|=|}
		\multirow{3}{*}{\rotatebox[origin=c]{90}{\parbox[c]{1.2cm}{\centering \textbf{Gao et al. \cite{SPIE14}}}}}
		 & 		CAVE           & 62.70    & 49.10        \\ \cline{2-4} 
		&Natural Scenes    & 4.02      & 52.35          \\ \cline{2-4} 
		&Cuprite           & 34.36     & 64.34  \\ \cline{1-4} 
                         
	\end{tabular}
\label{tab1}
\end{center}
\end{table}
\egroup
The results for different QP values can be found in Table~\ref{tab1}. While being more efficient at lower bitrates, the proposed method achieves significant BD rate savings for all evaluated data sets. On average, savings of 46~\% and 82~\% are found compared to HEVC Intra for QP values from 0 to 17 and from 22 to 37, respectively. In comparison to Gao et al. \cite{SPIE14}, average BD rate savings of 34~\% and 55~\% are achieved. It is assumed that the compression performance of PRIBP improves for high QP values, because more information is discarded due to quantization. Hence, the data available in the spectral reference becomes more valuable for preserving quality at low bitrates.
\captionsetup[subfloat]{labelformat=empty}
\begin{figure*}[htbp]
	\centering
\subfloat[Original: band 31 of scene "cloth\_ms" (CAVE).\label{4h}]{
	\includegraphics[width=0.3\textwidth]{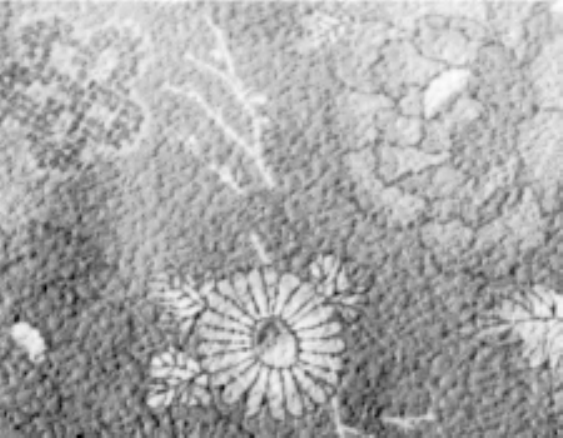}
}
\hfill
\subfloat[Original: band 18 of scene 5 \cite{Foster06}.\label{4i}]{
	\includegraphics[width=0.3\textwidth]{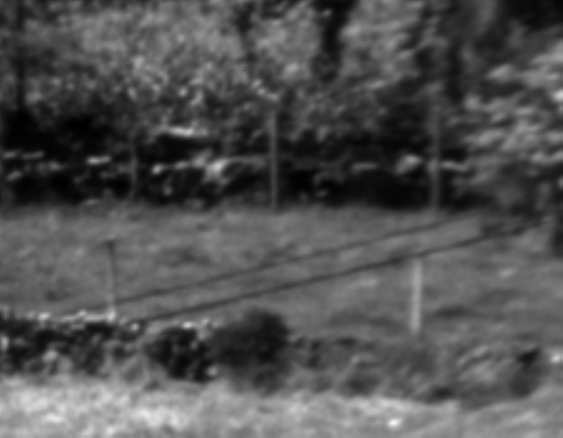}
}
\hfill
\subfloat[Original: band 13 of Cuprite scene (AVIRIS).\label{4j}]{
	\includegraphics[width=0.3\textwidth]{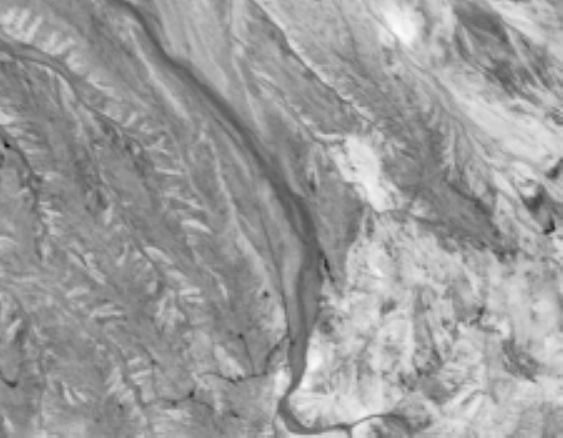}
}	
\hfill
	
\subfloat[Gao et al. \cite{SPIE14}:  29.70 dB, 0.061 bpp (QP=37)\label{4d}]{
	\includegraphics[width=0.3\textwidth]{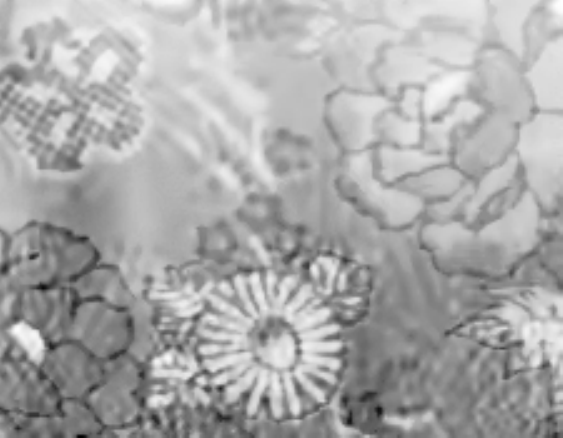}
}
\hfill
\subfloat[Gao et al. \cite{SPIE14}:  37.63 dB, 0.006 bpp (QP=37)\label{4e}]{
	\includegraphics[width=0.3\textwidth]{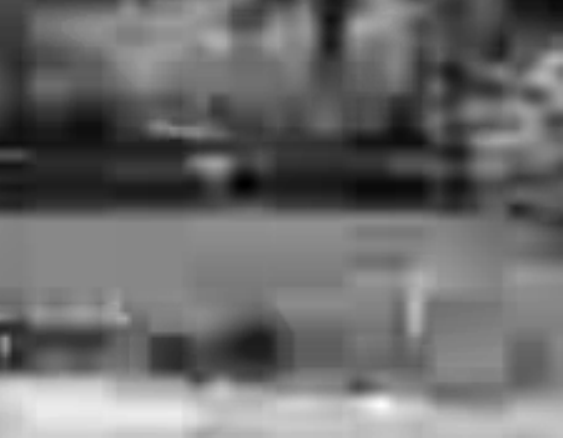}
}
\hfill
\subfloat[Gao et al. \cite{SPIE14}: 31.89 dB, 0.116 bpp (QP=37)\label{4f}]{
	\includegraphics[width=0.3\textwidth]{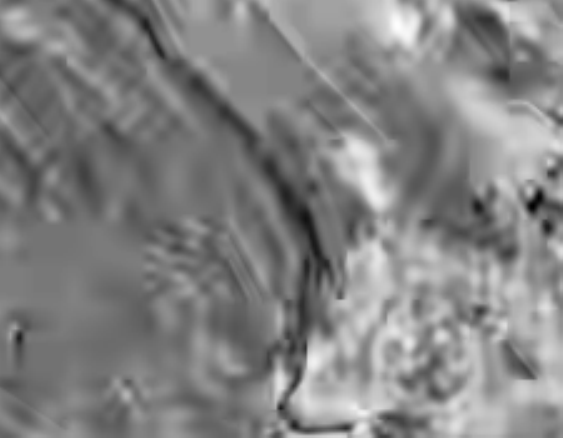}
}	
\hfill
	\subfloat[\textbf{Proposed PRIBP:  32.67 dB, 0.057 bpp (QP=33)}\label{4a}]{
		\includegraphics[width=0.3\textwidth]{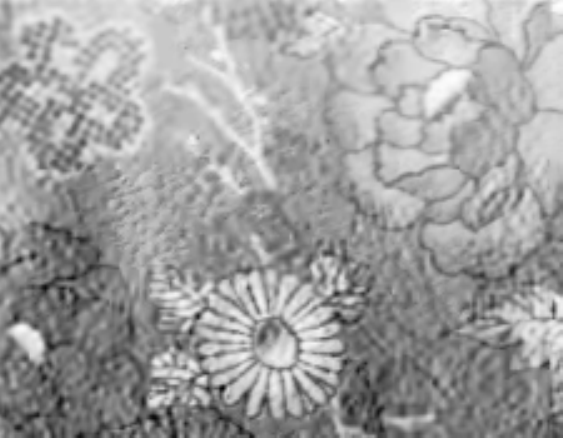}
	}
\hfill
	\subfloat[\textbf{Proposed PRIBP:  41.36 dB, 0.006 bpp (QP=34)}\label{4b}]{
	\includegraphics[width=0.3\textwidth]{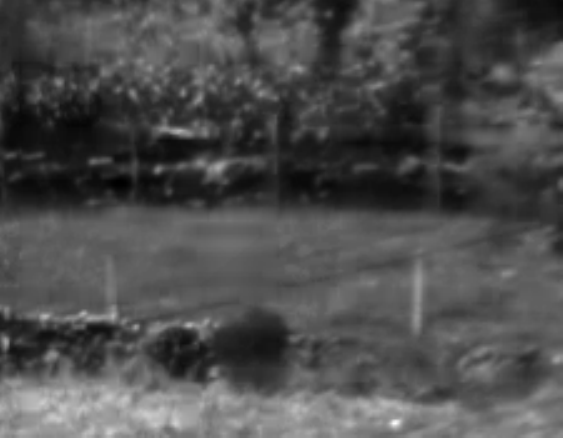}
}
\hfill
	\subfloat[\textbf{Proposed PRIBP: 48.39 dB, 0.096 bpp (QP=14)}\label{4c}]{
		\includegraphics[width=0.3\textwidth]{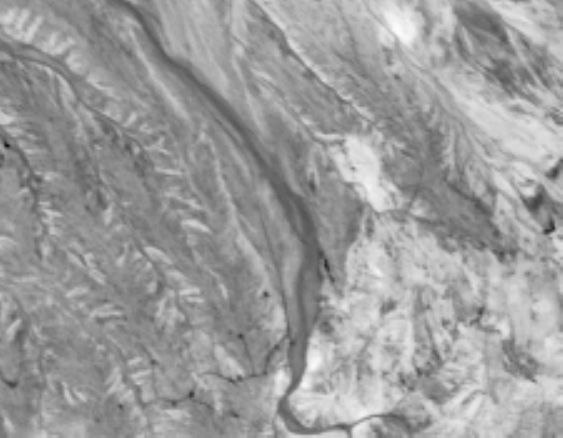}
	}
	\caption{Performance of exemplary images for every investigated data set in terms of PSNR in dB and bitrate in bits per pixel (bpp). The images show 16 bit reflectance data of different bands in comparison to the decoded versions.}
	\label{fig4}
\end{figure*}
Since the computation of the prediction signal in HEVC Intra only requires copying pixel values, the computational complexity for the inter-band prediction involving parameter estimation for two of the five additional modes is higher.
As a result, the average run time increases from 0.867 to 1.401~s for encoding and from 0.018 to 0.023~s at the decoder, whereby no optimizations for speed were made.

Fig.~\ref{fig4} provides visual examples of the performance on exemplary images of each data set. Clearly, PRIBP is capable of preserving sharp, complicated details, whereas Gao et al. \cite{SPIE14} blurs them at comparable bitrates and is not able to prevent blocking artifacts.

\section{Conclusion}\label{sec5}
In this contribution, a novel efficient multispectral compression scheme based on HEVC was introduced. By extending intra prediction with a spectral backward predictor based on linear regression modeling, the proposed method is able to exploit spatial as well as spectral redundancies without requiring additional side information for decoding. The conducted evaluation on different data sets demonstrates that the proposed method outperforms state-of-the-art methods. Compared to HEVC Intra, average BD rate savings of 46~\% and 82~\% were achieved for the QP values of 0-17 and 22-37, respectively. In comparison to Gao et al. \cite{SPIE14}, savings of 34~\% and 55~\% were found. These significant rate reductions were accomplished for a variety of image contents, especially for images containing complicated details. 

Furthermore, the proposed method is directly applicable to compression of multispectral videos as well.
The coding order can be extended to video in two possible ways. The first option is to code the first three bands of the video independently as a whole and subsequently ordering the remaining video bands based on the SSIM of the first frame. The second option is to apply the coding order for each frame individually, resulting in more complexity but possibly higher rate savings.

Like the state of the art, the proposed method is based on HEVC but can be transferred to Versatile Video Coding, which will be investigated in future work.

\bibliographystyle{IEEEtran}
\bibliography{bib_multispectral}

\end{document}

%% file: figures/cave_complete.tex
\begin{tikzpicture}

\definecolor{color0}{rgb}{0.8,0.2,0.2}
\definecolor{color1}{rgb}{0.15,0.7,0.15}
\definecolor{color2}{rgb}{0.216,0.255,1}

\begin{axis}[
axis line style={black},
legend cell align={left},
legend entries={{PRIBP},{Gao et al. \cite{SPIE14}},{HEVC Intra},{JPEG 2000}},
legend style={at={(1.001,0.33)}, anchor=north east,draw=black, fill=white},
tick align=outside,
tick pos=left,
x grid style={dashed, white!80.0!black},
xlabel={Bitrate [bpppb]},
xmajorgrids,
xmin=-0.01, xmax=0.44,
y grid style={dashed, white!80.0!black},
ylabel={PSNR [dB]},
ymajorgrids,
ymin=38.5984502253606, ymax=51.5
]
\addplot [thick, color0, mark=o, mark size=1, mark options={solid}]
table [row sep=\\]{%
0.160141981565035 51.0795633263221 \\
0.0842422216366499 48.3103862980769 \\
0.0463794194735013 45.6321237229567 \\
0.0255622007907965 42.8899309795673 \\
0.0146087744297125 40.1925807091346 \\
};

\addplot [thin, color1, mark=diamond, mark size=1, mark options={solid}]
table [row sep=\\]{%
0.235666201664851	50.6616391676683 \\
0.132421187865428	47.7959147836538 \\
0.0768067653362568	45.0029907902644 \\
0.0442512463300656	42.0787853064904 \\
0.0256824004344451	39.1046284254808 \\
};
\addplot [thin, color2, mark=triangle, mark size=1.5, mark options={solid,rotate=180}]
table [row sep=\\]{%
0.417892639453594	50.9834033503606 \\
0.253378819196652	48.1568735576923 \\
0.156726653759296	45.3678234825721 \\
0.0953813577309633	42.4098022536058 \\
0.0563245064172989	39.3496052283654 \\
};
\addplot [thin, black, mark=square, mark size=0.8, mark options={solid}]
table [row sep=\\]{%
0.410703854683118	46.7662023182763 \\
0.323312820532383	45.426425018403 \\
0.226816813151042	43.4823192124757 \\
0.174715298872728	42.0616807539052 \\
0.151954907637376	41.2965056821804 \\
};

\end{axis}

\end{tikzpicture}

%% file: figures/scene5+6.tex
\begin{tikzpicture}

\definecolor{color0}{rgb}{0.8,0.2,0.2}
\definecolor{color1}{rgb}{0.15,0.7,0.15}
\definecolor{color2}{rgb}{0.216,0.255,1}

\begin{axis}[
axis line style={black},
legend cell align={left},
legend entries={{PRIBP},{Gao et al. \cite{SPIE14}},{HEVC Intra},{JPEG 2000}},
legend style={at={(1.001,0.33)}, anchor=north east,draw=black, fill=white},
tick align=outside,
tick pos=left,
x grid style={dashed, white!80.0!black},
xlabel={Bitrate [bpppb]},
xmajorgrids,
xmin=-0.01, xmax=0.53,
y grid style={dashed, white!80.0!black},
ylabel={PSNR [dB]},
ymajorgrids,
ymin=36.5972925168011, ymax=49.3
]
\addplot [thick, color0, mark=o, mark size=1, mark options={solid}]
table [row sep=\\]{%
	0.277240366232229	48.8689466599462 \\
	0.106364894882152	46.3705129569893 \\
	0.0423299926689674	44.2760215053763 \\
	0.0170753922263001	42.3630814919355 \\
	0.00777576588419977	40.645873608871 \\
};

\addplot [thin, color1, mark=diamond, mark size=1, mark options={solid}]
table [row sep=\\]{%
	0.304258016225613	48.7754509610215 \\
	0.125800312844804	45.96772 \\
	0.054812976916373	43.2963750537634 \\
	0.0265144959847471	40.6011979973118 \\
	0.0151304177691522	37.9894471034946 \\
};

\addplot [thin, color2, mark=triangle, mark size=1.5, mark options={solid,rotate=180}]
table [row sep=\\]{%
	0.500156671670629	49.0268596706989 \\
	0.285433427869517	46.2575279569893 \\
	0.161603862379381	43.5243074193548 \\
	0.0887179339145771	40.8029438037634 \\
	0.0478312625870095	38.1829677486559 \\
};

\addplot [thin, black, mark=square, mark size=0.8, mark options={solid}]
table [row sep=\\]{%
0.502612821093906	47.8107641234671 \\
0.382921640843282	46.4053808464183 \\
0.281973644785613	44.9099134120008 \\
0.208615503308851	43.4453986860789 \\
0.165561941902327	42.3050705630228 \\
};

\end{axis}

\end{tikzpicture}

%% file: figures/cuprite.tex
\begin{tikzpicture}

\definecolor{color0}{rgb}{0.8,0.2,0.2}
\definecolor{color1}{rgb}{0.15,0.7,0.15}
\definecolor{color2}{rgb}{0.216,0.255,1}

\begin{axis}[
axis line style={black},
legend cell align={left},
legend entries={{PRIBP},{Gao et al. \cite{SPIE14}},{HEVC Intra},{JPEG 2000}},
legend style={at={(1.001,0.33)}, anchor=north east,draw=black, fill=white},
tick align=outside,
tick pos=left,
x grid style={dashed, white!80.0!black},
xlabel={Bitrate [bpppb]},
xmajorgrids,
xmin=-0.0350623683975372, xmax=2.33,
y grid style={dashed, white!80.0!black},
ylabel={PSNR [dB]},
ymajorgrids,
ymin=30.4993961162511, ymax=46.5
]
\addplot [thick, color0, mark=o, mark size=1, mark options={solid}]
table [row sep=\\]{%
0.535840255270145	45.6360325980392 \\
0.283251331645597	42.0822354779412 \\
0.155185296603565	39.2653555147059 \\
0.0870945434780272	36.8501547794118 \\
0.0503434849474329	34.7937005514706 \\
};
\addplot [thin, color1, mark=diamond, mark size=1, mark options={solid}]
table [row sep=\\]{%
1.0064951316327	45.5026629901961 \\
0.545932030981252	41.0877526348039 \\
0.288533318686576	37.3999976715686 \\
0.149504974858639	34.3164709558824 \\
0.0778898036393834	31.7026250612745 \\
};
\addplot [thin, color2, mark=triangle, mark size=1.5, mark options={solid,rotate=180}]
table [row sep=\\]{%
2.03344810546203	46.1215036764706 \\
1.23929304366346	41.5213227328431 \\
0.676131126107241	37.6571055147059 \\
0.337824571817488	34.4991822303922 \\
0.158245200835268	31.884740747549 \\
};
\addplot [thin, black, mark=square, mark size=0.8, mark options={solid}]
table [row sep=\\]{%
2.28191253309531	45.2347576997308 \\
1.33097733973845	40.2148538212373 \\
0.725270153040757	36.7407544906715 \\
0.379174397264121	34.1439700800774 \\
0.27842436216303	33.0469080525542 \\
0.164078796935173	31.5976385922746 \\
};

\end{axis}

\end{tikzpicture}

%% file: multispectral_compression_hevc.bbl
\begin{thebibliography}{10}
\providecommand{\url}[1]{#1}
\csname url@samestyle\endcsname
\providecommand{\newblock}{\relax}
\providecommand{\bibinfo}[2]{#2}
\providecommand{\BIBentrySTDinterwordspacing}{\spaceskip=0pt\relax}
\providecommand{\BIBentryALTinterwordstretchfactor}{4}
\providecommand{\BIBentryALTinterwordspacing}{\spaceskip=\fontdimen2\font plus
\BIBentryALTinterwordstretchfactor\fontdimen3\font minus
  \fontdimen4\font\relax}
\providecommand{\BIBforeignlanguage}[2]{{%
\expandafter\ifx\csname l@#1\endcsname\relax
\typeout{** WARNING: IEEEtran.bst: No hyphenation pattern has been}%
\typeout{** loaded for the language `#1'. Using the pattern for}%
\typeout{** the default language instead.}%
\else
\language=\csname l@#1\endcsname
\fi
#2}}
\providecommand{\BIBdecl}{\relax}
\BIBdecl

\bibitem{tretter05}
D.~Tretter, N.~Memon, and C.~Bouman, \emph{\BIBforeignlanguage{English
  (US)}{Multispectral Image Coding}}.\hskip 1em plus 0.5em minus 0.4em\relax
  Elsevier Inc., Dec. 2005, pp. 747--760.

\bibitem{huang10}
Y.~Huang, S.~Thomson, Y.~Lan, and S.~Maas, ``Multispectral imaging systems for
  airborne remote sensing to support agricultural production management,''
  \emph{International Journal of Agricultural and Biological Engineering},
  vol.~3, Jan. 2010.

\bibitem{takumi17}
K.~Takumi, K.~Watanabe, Q.~Ha, A.~Tejero-De-Pablos, Y.~Ushiku, and T.~Harada,
  ``Multispectral object detection for autonomous vehicles,'' in \emph{Proc.
  Thematic Workshops of ACM Multimedia}, Oct. 2017, p. 35–43.

\bibitem{Kim16}
S.~Kim, D.~Cho, J.~Kim, M.~Kim, S.~Youn, J.~Jang, M.~Je, D.-H. Lee, B.~Lee,
  D.~Farkas, and J.~Hwang, ``Smartphone-based multispectral imaging: System
  development and potential for mobile skin diagnosis,'' \emph{Biomedical
  Optics Express}, vol.~7, pp. 5294--5307, Dec. 2016.

\bibitem{Salamati12}
N.~{Salamati}, Z.~{Sadeghipoor}, and S.~{Süsstrunk}, ``{Compression of
  multispectral images: Color (RGB) plus Near-Infrared (NIR)},'' in \emph{Proc.
  IEEE International Workshop on Multimedia Signal Processing (MMSP)}, Sep.
  2012, pp. 65--70.

\bibitem{SPIE14}
F.~Gao, X.~Ji, C.~Yan, and Q.~Dai, ``{Compression of multispectral image using
  HEVC},'' in \emph{Proc. Optoelectronic Imaging and Multimedia Technology
  III}, vol. 9273, International Society for Optics and Photonics.\hskip 1em
  plus 0.5em minus 0.4em\relax SPIE, Oct. 2014, pp. 592 -- 601.

\bibitem{nguyen15}
T.~{Nguyen} and D.~{Marpe}, ``Objective performance evaluation of the {HEVC}
  main still picture profile,'' \emph{IEEE Transactions on Circuits and Systems
  for Video Technology}, vol.~25, no.~5, pp. 790--797, May 2015.

\bibitem{ryan97}
M.~J. {Ryan} and J.~F. {Arnold}, ``The lossless compression of {AVIRIS} images
  by vector quantization,'' \emph{IEEE Transactions on Geoscience and Remote
  Sensing}, vol.~35, no.~3, pp. 546--550, 1997.

\bibitem{Lin10}
C.~{Lin} and Y.~{Hwang}, ``An efficient lossless compression scheme for
  hyperspectral images using two-stage prediction,'' \emph{IEEE Geoscience and
  Remote Sensing Letters}, vol.~7, no.~3, pp. 558--562, Apr. 2010.

\bibitem{tang03}
X.~Tang, W.~A. Pearlman, and J.~W. Modestino, ``{Hyperspectral image
  compression using three-dimensional wavelet coding},'' in \emph{Proc. Image
  and Video Communications and Processing}, vol. 5022, International Society
  for Optics and Photonics.\hskip 1em plus 0.5em minus 0.4em\relax SPIE, May
  2003, pp. 1037 -- 1047.

\bibitem{wang09}
L.~{Wang}, J.~{Wu}, L.~{Jiao}, and G.~{Shi}, ``Lossy-to-lossless hyperspectral
  image compression based on multiplierless reversible integer {TDLT}/{KLT},''
  \emph{IEEE Geoscience and Remote Sensing Letters}, vol.~6, no.~3, pp.
  587--591, Jul. 2009.

\bibitem{CCSD}
{\relax Consultative Committee for Space Data Systems}, ``Low-complexity
  lossless and near-lossless multispectral and hyperspectral image
  compression,'' Feb. 2019, {R}ecommended Standard CCSDS 123.0-B-2.

\bibitem{CCSD_spectral}
------, ``Spectral preprocessing transform for multispectral and hyperspectral
  image compression,'' Sep. 2017, {R}ecommended Standard CCSDS 122.1-B-1.

\bibitem{Du07}
Q.~{Du} and J.~E. {Fowler}, ``Hyperspectral image compression using {JPEG}2000
  and {P}rincipal {C}omponent {A}nalysis,'' \emph{IEEE Geoscience and Remote
  Sensing Letters}, vol.~4, no.~2, pp. 201--205, Apr. 2007.

\bibitem{penna07}
B.~{Penna}, T.~{Tillo}, E.~{Magli}, and G.~{Olmo}, ``Transform coding
  techniques for lossy hyperspectral data compression,'' \emph{IEEE
  Transactions on Geoscience and Remote Sensing}, vol.~45, no.~5, pp.
  1408--1421, May 2007.

\bibitem{B_scones_2018}
D.~Báscones, C.~González, and D.~Mozos, ``Hyperspectral image compression
  using vector quantization, {PCA} and {JPEG}2000,'' \emph{Remote Sensing},
  vol.~10, no.~6, p. 907, Jun. 2018.

\bibitem{Dusselaar15}
R.~{Dusselaar}, M.~{Paul}, and T.~{Bossomaier}, ``{Hyperspectral image coding
  using Spectral Prediction Modelling in HEVC coding framework},'' in
  \emph{Proc. International Conference on Image and Vision Computing New
  Zealand (IVCNZ)}, Nov 2015, pp. 1--6.

\bibitem{Paul16}
M.~Paul, R.~Xiao, J.~Gao, and T.~Bossomaier, ``Reflectance prediction modelling
  for residual-based hyperspectral image coding,'' \emph{PLOS ONE}, vol.~11,
  no.~10, pp. 1--16, Oct. 2016.

\bibitem{sullivan}
G.~J. {Sullivan}, J.~{Ohm}, W.~{Han}, and T.~{Wiegand}, ``{Overview of the High
  Efficiency Video Coding (HEVC) Standard},'' \emph{IEEE Transactions on
  Circuits and Systems for Video Technology}, vol.~22, no.~12, pp. 1649--1668,
  Dec. 2012.

\bibitem{Lainema12}
J.~{Lainema}, F.~{Bossen}, W.~{Han}, J.~{Min}, and K.~{Ugur}, ``{Intra Coding
  of the HEVC Standard},'' \emph{IEEE Transactions on Circuits and Systems for
  Video Technology}, vol.~22, no.~12, pp. 1792--1801, Dec. 2012.

\bibitem{Genser19}
N.~{Genser}, J.~{Seiler}, and A.~{Kaup}, ``Joint regression modeling and sparse
  spatial refinement for high-quality reconstruction of distorted color
  images,'' in \emph{Proc. IEEE International Conference on Image Processing
  (ICIP)}, Sep. 2019, pp. 3262--3266.

\bibitem{Wang04}
{Zhou Wang}, A.~C. {Bovik}, H.~R. {Sheikh}, and E.~P. {Simoncelli}, ``Image
  quality assessment: from error visibility to structural similarity,''
  \emph{IEEE Transactions on Image Processing}, vol.~13, no.~4, pp. 600--612,
  Apr. 2004.

\bibitem{Foster06}
D.~H. Foster, K.~Amano, S.~M.~C. Nascimento, and M.~J. Foster, ``Frequency of
  metamerism in natural scenes,'' \emph{Journal of the Optical Society of
  America A}, vol.~23, no.~10, pp. 2359--2372, Oct. 2006.

\bibitem{qps}
F.~Bossen, ``{Common test conditions and software reference configurations},''
  Jan. 2013, {document JCTVC-L1100 of JCT-VC, Geneva, CH}.

\bibitem{bdrate}
G.~Bj{\o}ntegaard, ``{Calculation of Average PSNR Differences between RD
  Curves},'' Apr. 2001, {document VCEG-M33, Austin, TX, USA}.

\end{thebibliography}
